\documentclass[preprint,aps,prd,showpacs,superscriptaddress,nofootinbib]{revtex4}

\usepackage{amsfonts}
\usepackage{mathrsfs}
\usepackage{graphicx}
\usepackage{amsmath}
\usepackage{amssymb}
\usepackage{bm}
\usepackage{bbm}

\def\OMIT#1{}

\newcommand{\nn}{\nonumber}

\newcommand{\beq}{\begin{equation}}
\newcommand{\eeq}{\end{equation}}
\newcommand{\bqa}{\begin{eqnarray}}
\newcommand{\eqa}{\end{eqnarray}}

\begin{document}

\title{\mbox{}\\[10pt]
Observation prospects of leptonic and Dalitz decays of pseudoscalar
quarkonia}

\author{Yu Jia}
\affiliation{Institute of High Energy Physics, Chinese Academy of
Sciences, Beijing 100049, China\vspace{0.2cm}}
\affiliation{Theoretical Physics Center for Science Facilities,
Chinese Academy of Sciences, Beijing 100049, China\vspace{0.2cm}}

\author{Wen-Long Sang}
\affiliation{Key Laboratory of Frontiers in Theoretical Physics,
Institute of Theoretical Physics, Chinese Academy of Sciences,
Beijing 100190, China\vspace{0.2cm}}

\date{\today}
\begin{abstract}
Two types of pseudoscalar quarkonium electromagnetic decay
processes, i.e. decay to a lepton pair, and to a lepton pair plus a
photon (Dalitz decay), are analyzed at the leading order in NRQCD
expansion. The former type of processes, highly suppressed in the
Standard Model, have been hoped to act as the sensitive probes of
the possible new physics. The latter type of processes generally
possess much greater decay rates than the former, owing to several
conspiring factors. The recently launched BES-III program, with
$10^8$ $\eta_c$ samples to be anticipated in the coming years, may
be able to observe the Dalitz decays $\eta_c\to e^+e^-\gamma$ and
$\eta_c\to \mu^+\mu^-\gamma$, which have branching ratios of order
$10^{-6}$. When the radiated photon becomes very soft, the Dalitz
decay events will be experimentally tagged as the exclusive lepton
pair events. It is found that, those quasi-two-body events that
arise from $\eta_c\to e^+ e^-\gamma$ with photon energy less than
the minimum sensitivity of the electromagnetic calorimeter, can
vastly outnumber the literal $\eta_c\to e^+ e^-$ events, however
this amplification is still not dramatic enough for the BES-III
experiment to establish these events. Consequently, the expectation
of looking for new physics signature in the $\eta_c\to l^+l^-$
channel is obscured, unless the contamination from $\eta_c \to
l^+l^-\gamma$ has been taken into account carefully.
\end{abstract}

\maketitle

\section{Introduction}

Charge-neutral pseudoscalar meson decay to a lepton pair, $P\to
l^+l^-$, has long been an interesting topic. In particular, for $P$
to be a light pseudoscalar meson ($\pi^0$, $\eta$, $\eta'\cdots$),
extensive theoretical and experimental efforts have been conducted
since Drell initiated the study of $\pi^0\to e^+e^-$ in
1959~\cite{Drell:1959,Berman:1960,Young:1967zz,Bergstrom:1982zq,
Bergstrom:1983ay,Savage:1992ac,Ametller:1993we,GomezDumm:1998gw,Dorokhov:2007bd}.
Studies of these decays can offer insights into the nonperturbative
structure of the pseudoscalar mesons, in particular, help one to
glean more knowledge about the $P\gamma^*\gamma^*$ transition form
factor. This type of electromagnetic decay processes are suppressed
by two additional powers of $\alpha$ with respect to
$P\to\gamma\gamma$, and also penalized by helicity conservation. As
a consequence, the decay probabilities are generally very tiny
within the Standard Model (SM), rendering experimental detection
rather challenging. On the other hand, the rareness of these decay
processes may turn into a virtue, that is, they might be utilized as
the sensitive probes of possible new interactions beyond SM.

Another intimately related type of electromagnetic decays are $P\to
l^+l^-\gamma$. These kinds of pseudoscalar meson decay processes are
of use to extract the information about the $P\gamma^*\gamma$ form
factor. Despite radiating off an extra photon, these bremsstrahlung
leptonic decays in general occur much more copiously than $P\to
l^+l^-$. For example, the so-called Dalitz decay
process~\cite{Dalitz:1951aj}, $\pi^0\to e^+e^-\gamma$, which has
been observed decades ago, has a branching ratio of
$(1.198\pm0.032)\%$~\cite{Amsler:2008zzb}. This is more than five
orders of magnitude greater than the branching fraction of $\pi^0\to
e^+e^-$, $(7.48\pm0.38)\times 10^{-8}$, which was recently measured
in the KTeV E799-II experiment at Fermilab~\cite{Abouzaid:2006kk}.
This striking disparity can be attributed to a number of facts, that
such bremsstrahlung leptonic decay processes are suppressed with
respect to $P\to\gamma\gamma$ by only one additional power of
$\alpha$, suffer no helicity suppression, and also enjoy collinear
enhancement brought in by photon fragmentation to a lepton pair.

In this work, we aim to investigate both types of leptonic decay
processes for $P$ to be a pseudoscalar heavy quarkonium state, i.e.,
$\eta_Q \to l^+l^-$ and $\eta_Q \to l^+l^-\gamma$
($Q=c,b$)~\footnote{In literature, the term {\it Dalitz decay} has
been specifically reserved for $\pi^0\to
e^+e^-\gamma$~\cite{Dalitz:1951aj}. In this work we generalize the
use of this term, i.e. we also use it to refer to any of the
$\eta_Q\to l^+l^-\gamma$ process.}. One strong incentive stems from
the experimental side. For instance, the recently launched BES-III
experiment, plans to accumulate an unprecedentedly large data set of
charmonia, {\it e.g.}, $10^{10}$ $J/\psi$ and $3\times 10^9$
$\psi(2S)$ in the coming years~\cite{Asner:2008nq}. An enormous
number of $\eta_c$ and $\eta_c(2S)$ are expected to be produced via
the radiative transitions from these $J/\psi$ or $\psi(2S)$ samples.
Analogously, the scheduled Super \textsc{Belle} experiment, will
also be capable of collecting a tremendous number of $\eta_b$
samples. Further, an even larger data set of pseudoscalar quarkonia
are expected to be produced at the CERN Large Hadron Collider (LHC),
though the copious backgrounds in hadron machine renders the
detection of such rare decays rather challenging. In any event, it
seems, at the current time, not of only academic interests to assess
the observation potentials of these rare electromagnetic decay
processes in the forthcoming experiments.

From a theoretical perspective, it is also worthwhile to study these
rare electromagnetic decays of pseudoscalar quarkonia, to enrich our
knowledge about heavy quark physics. A heavy quarkonium state, being
a heavy-quark heavy-antiquark pair tightly bound via the strong
interaction, is the best understood among all types of hadrons. In
sharp contrast to light meson decay, which must be analyzed by some
nonperturbative tools, quarkonium decay can be accommodated in the
perturbative QCD framework, owing to the condition $m_Q\gg
\Lambda_{QCD}$. Indeed, annihilation decays of heavy quarkonium,
especially the electromagnetic ones that we plan to investigate, can
be systematically tackled by the modern effective-field-theory
formalism, the nonrelativistic QCD (NRQCD) factorization
approach~\cite{Bodwin:1994jh,Brambilla:2004jw}. In a quarkonium
electromagnetic decay process, this factorization approach allows
one to systematically separate the hard quantum fluctuation of order
heavy quark mass $m$ from the low-energy contributions of order $mv$
or smaller, where $v$ signifies the typical velocity of $Q$ or
$\overline{Q}$ in a quarkonium. Empirically, $v^2 \approx 0.3$ for
charmonium, and $0.1$ for bottomonium.

We note that, both types of leptonic decay processes of $\eta_Q$
have already been partly investigated by different authors. For
instance, $\eta_Q\to l^+l^-$ has been studied in
Ref.~\cite{Bergstrom:1982zq,YJZhang:2007:thesis,Yang:2009kq}, and
the Dalitz decay $\eta_Q\to l^+l^-\gamma$ was considered in
\cite{Bergstrom:1980fr,DiSalvo:2000ec}. Nevertheless, a
comprehensive analysis based on the NRQCD factorization approach is
still lacking~\footnote{Ref.~\cite{Bergstrom:1982zq} studies the
$\eta_Q\to l^+l^-$ process within a bound state quark model, which
nevertheless may be viewed as a primitive version of the NRQCD
approach. Ref.~\cite{YJZhang:2007:thesis} employs the NRQCD
factorization explicitly, but only studies a few processes using
numerical recipe. Both of works completely neglect the weak
interaction contribution, which turns out to be inadequate for
$\eta_b$ decay.}. For this reason, we feel that it might be
rewarding to revisit these two processes from this angle. We will
work at the leading order in NRQCD expansion only. However, the
systematics of this approach renders future implementation of
higher-order corrections possible.

We summarize the main outcome of this work. We confirm the
analytic expression for the electromagnetic contribution to the
decay $\eta_Q\to l^+l^-$, which was first reported in
Ref.~\cite{Bergstrom:1982zq}. We also include the often-omitted weak
interaction contribution arising from $Z^0$ exchange. It is found
that, although the weak interaction plays a negligible role in the
leptonic decay of $\eta_c$, its effect can become important in
$\eta_b$ decay, especially for $\eta_b\to
\tau^+\tau^-$~\footnote{The weak interaction effect for $\eta_c\to
l^+l^-$ has also been considered in \cite{Yang:2009kq}.}. Despite
this, the net SM predictions to the decay rates are still too
suppressed for these processes to be observed experimentally in the
foreseeable future.

We also perform a comprehensive study of numerous pseudoscalar
quarkonium Dalitz decay processes, and verify that such decays
generally possess much more enhanced decay probability than
$\eta_Q\to l^+l^-$. In particular, it is found that $\eta_c\to e^+
e^-\gamma,\:\mu^+\mu^-\gamma$, with branching fractions of order
$10^{-6}$, might have bright prospect to be observed at the BES-III
experiment. We also analyze the energy distributions of lepton and
photon in these Dalitz decay processes. We hope future measurements
of these energy spectra can test our predictions critically.

We emphasize that any realistic electromagnetic calorimeter is
limited by the finite sensitivity to detect soft photons. This
indicates that, the Dalitz decay event $\eta_Q \to l^+l^-\gamma$
will fake the literal $\eta_Q\to l^+l^-$ event, if the emitted
photon is too soft to be registered by the electromagnetic
calorimeter. Therefore, the hope of seeking new interaction beyond
SM in the $\eta_Q\to l^+l^-$ channel becomes obscured, unless the
contamination from the respective Dalitz decay is thoroughly
understood and incorporated in the analysis. Our study shows that,
the number of quasi-two-body events that arise from $\eta_c\to e^+
e^-\gamma$ with photon energy restricted to be less than 20 MeV, can
easily surpass that from $\eta_c\to e^+ e^-$ by two orders of
magnitude, but such enhancement is still not significant enough to
warrant the establishment of such events at BESIII experiment.

The remainder of the paper is distributed as follows. In
Section~\ref{P:lepton:pair}, we calculate the pseudoscalar quarkonia
decays to a lepton pair at the leading order in NRQCD expansion,
including both contributions from QED and weak interaction. We also
compare our results with the previous ones. In
Section~\ref{etaQ:Dalitz:decay}, we present a detailed calculation
of the pseudoscalar quarkonium Dalitz decay processes at the leading
order in NRQCD expansion. The inclusive energy spectra of the photon
and the lepton are presented, and the QED fragmentation function for
a photon to split into a lepton can be extracted. We also obtain a
succinct expression for the decay branching fraction integrated over
the full three-body phase space. In Section~\ref{phenomenology}, a
comprehensive numerical analysis for numerous processes of
pseudoscalar quarkonia decays to a lepton pair and the Dalitz decays
are made. Particular attention is paid to the interference pattern
between QED and the $Z^0$-exchange contributions to the former type
of processes. We assess the observation prospects of both types of
decay processes. We also investigate to which extent the Dalitz
decay events will fake the exclusive lepton pair events. In
Section~\ref{summary}, we summarize and present an outlook.

\section{Pseudoscalar quarkonium decay to a lepton pair}
\label{P:lepton:pair}

\begin{figure}[!htb]
\centerline{
\includegraphics[height=3.4cm]{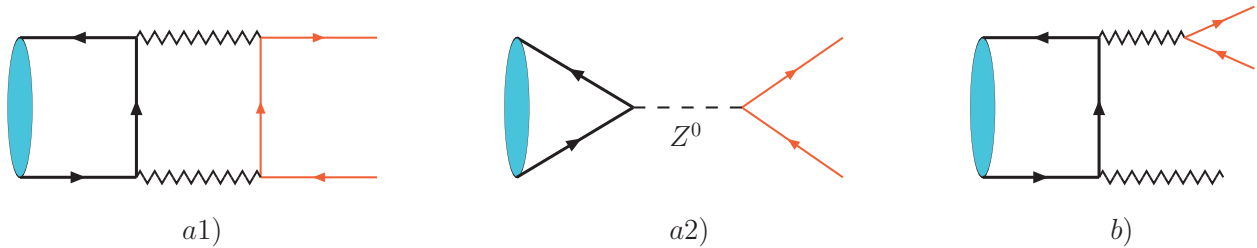}
} \caption{Lowest-order diagrams for the processes $\eta_Q\to l^+
l^-$ [$a$] and $\eta_Q\to l^+ l^-\gamma$ [$b$]. In the former
process, we represent the QED contribution by $a1)$, the weak
interaction contribution by $a2)$. The crossed diagrams for $a1)$
and $b)$ have been suppressed.}
\label{Feynman:diag}
\end{figure}

The purpose of this section is to derive the analytic expressions
for the amplitude of $\eta_Q\to l^- l^+$ in the lowest-order NRQCD
expansion, by taking only the SM interactions into consideration.
That is, we will consider the contributions from both
electromagnetic and weak interactions, and investigate their
interference pattern.

\subsection{The electromagnetic contribution to $\eta_Q\to l^- l^+$}

In this subsection we focus on the QED contribution to this process.
Unlike the leptonic decay of vector quarkonium such as $J/\psi$, a
pseudoscalar quarkonium cannot directly decay to a lepton pair
through annihilation into a virtual photon at tree level. At the
lowest order in electromagnetic and strong couplings, this process
proceeds through the one-loop QED box diagram, as depicted in
Fig.~\ref{Feynman:diag}a)~\footnote{The same annihilation type of
diagram, with the lepton replaced by the light quark and the photon
replaced by gluon, has been hypothesized to account for the
$\eta_c-\eta(\eta')$ mixing~\cite{Isgur:1975ib}.}.

The validity of the NRQCD approach rests upon one of the key
characteristics of quarkonium, that both of its constitutes move
non-relativistically in the quarkonium rest frame, so the quark
relative velocity can serve as a small expansion parameter of the
theory. For this reason, NRQCD would be a very poor framework to
describe light mesons such as $\pi$, $\eta$, $\eta'$, etc. Nowadays
the NRQCD approach has been accepted as the standard tool to analyze
heavy quarkonium decay and production processes. Since hard
reactions involving quarkonium necessarily probe the scale of order
heavy quark mass, by appealing to the asymptotic freedom of QCD,
NRQCD approach allows one to put the amplitude in a factorized form,
i.e., the sum of the products of perturbatively calculable
short-distance coefficients and nonperturbative but universal NRQCD
matrix elements.

The calculation of quarkonium decay at the LO in NRQCD expansion is
standard. We assume the quarkonium state composed of a heavy quark
$Q$ and its antiquark $\overline{Q}$, and abbreviate the
$Q\overline{Q}({}^1S_0^{(1)})$ state by $\eta_Q$. We assign the
momenta carried by $\eta_Q$, $l^-$, $l^+$ as $P$, $p_1$, $p_2$. At
the LO in velocity expansion, one can routinely obtain the amplitude
for $\eta_Q \to l^- l^+$ by first computing the amplitude for
$Q({P\over 2})\overline{Q}({P\over 2})\to l^-(p_1)l^+(p_2)$,
enforcing $Q$ and $\overline{Q}$ to carry equal momentum, then
projecting it onto the intended ${}^1S_0^{(1)}$ state.

There are totally two lowest-order QED diagrams for this process,
one of which is shown in Fig.~\ref{Feynman:diag}-$a1)$. The other
undrawn diagram can be obtained from it by reversing the fermionic
arrow in either the quark or the lepton line. By $C$-invariance,
both of diagrams yield the identical results. After some
straightforward calculation, we can express the electromagnetic
decay amplitude at the LO in NRQCD expansion as
\bqa
{\mathscr M}_{\rm EM}[\eta_Q \to l^- l^+] &=& 2\sqrt{2 N_c} e_Q^2
e_l^2 \alpha^2 {\psi_{\eta_Q}(0)\over m_Q^{5/2} } \, f\left({m_l^2\over m_Q^2}\right)
\,[ m_l \, \bar{u}(p_1)\gamma_5 v(p_2)],
\label{QED:ampl:etac:l+:l-}
\eqa
where $N_c=3$ is the number of colors. $e_l$ and $e_Q$ signify the
electric charges of charged lepton and heavy quark in units of $|e|$
($e_l=-1$, $e_c=2/3$ and $e_b=-1/3$), and $\alpha$ is the fine
structure constant. The nonperturbative factor $\psi_{\eta_c}(0)$,
is the wave function at the origin for the $\eta_Q$ state, which can
be identified with the LO NRQCD matrix element. $m_l$ and $m_Q$
denote the masses of lepton and quark, respectively. At the current
level of accuracy, it is legitimate to treat $m_Q$ and
$M_{\eta_Q}/2$ interchangeably. The dimensionless function $f$
encodes the effect of loop contribution, and is normalized in such a
way that it depends on the ratio of lepton mass to quark mass at
most logarithmically.

Some remarks on the traits of this process are in order. Because the
lepton pair must form a $^1S_0$ state to conserve angular momentum,
a $\gamma_5$ is expected to be sandwiched between the leptonic
spinors in the decay amplitude. Indeed, with the aid of Dirac
equation, one easily verifies that the bispinor
$\bar{u}(p_1)\gamma_5 v(p_2)$ exhausts all the possible Lorentz
structures. It is well known that leptonic decays of pseudoscalar
meson suffer from the so-called {\it helicity suppression}, and
would be strictly forbidden when lepton mass set to zero, hence
there should be an explicit factor of lepton mass appearing in the
amplitude~\footnote{We stress this phenomenon not only pertains to
pseudoscalar meson leptonic decay. The decay $\chi_{c0}\to e^+e^-$
should also be forbidden in the zero-$m_e$ limit, again due to the
conflict between angular momentum conservation and helicity
conservation in massless QED.}. Note that the $f$ function diverges
with $m_l$ only logarithmically, so equation
(\ref{QED:ampl:etac:l+:l-}) is compatible with the helicity
suppression mechanism.

\begin{figure}[!tb]
\centerline{
\includegraphics[angle=270,width=15 cm]{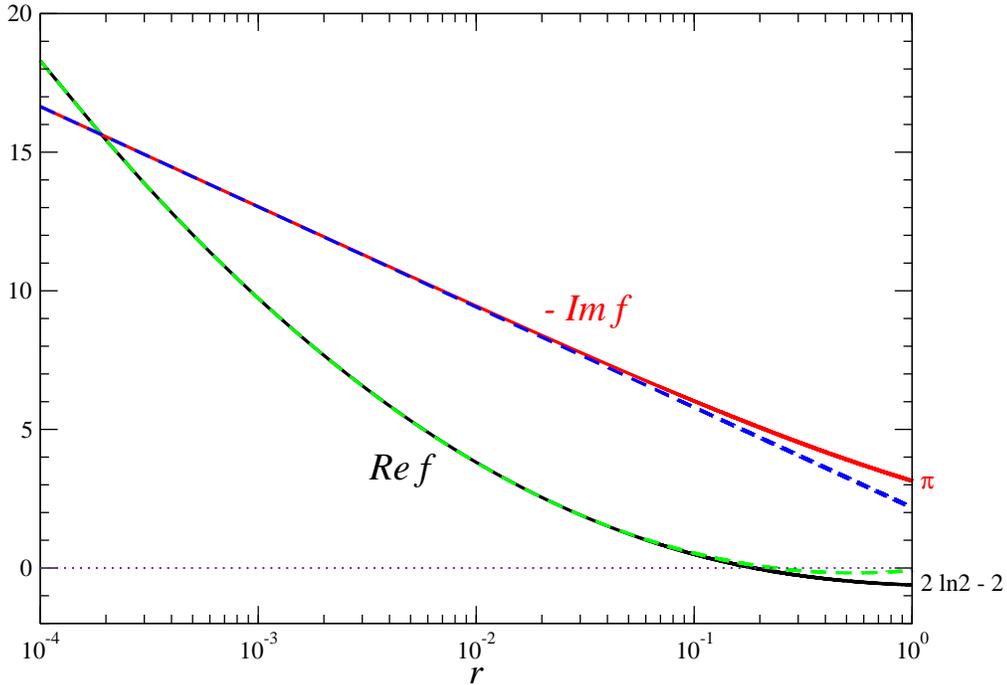}
 }
 \caption{The profile of the function $f(r)$. Solid lines represent the
exact results given in (\ref{analy:f:function}), and the dashed ones
represent the asymptotic expressions as given in
(\ref{Re:f:asymptotic}) and (\ref{Im:f:asymptotic}).
\label{shape:of:f}}
\end{figure}

It turns out that the evaluation of the box diagram can be reduced
to evaluating a three-point one-loop integral. The encountered loop
integrals are both ultraviolet and infrared finite, hence may be
directly computed at four spacetime dimension. After some efforts,
we can obtain the closed form for the $f$ function:
\bqa
f(r) &=& {1\over \beta}\left[{1\over 4} \ln^2 \left({1+\beta\over
1-\beta} \right)  - \ln \left({1+\beta\over 1-\beta} \right) +
{\pi^2\over 12} + {\rm Li}_2\left(-{1-\beta\over 1+\beta}\right) -
{i\pi\over 2} \ln \left({1+\beta\over 1-\beta} \right) \right],
\label{analy:f:function}
\eqa
where ${\rm Li}_2$ is the dilogarithm. We have introduced $r\equiv
{4 m_l^2\over M^2_{\eta_Q}}={m_l^2\over m_Q^2}$, and $\beta \equiv
\sqrt{1-r}$ is the velocity of the outgoing lepton in the $\eta_Q$
rest frame. The shape of this function is shown in
Fig.~\ref{shape:of:f}. Note Eq.~(\ref{analy:f:function}) is
identical to the $R$ function given in equation~(12) of
\cite{Bergstrom:1982zq}, once the Spence function $\Phi(x)$ there is
identified with $-{\rm Li}_2(-x)$. It may be worth pointing out that,
in order to reach the compact expression given in
(\ref{analy:f:function}), we have made use of the following somewhat
inapparent relation~\footnote{This relation can be proven with the
help of the identity ${\rm Li}_2({x\over 2})={\rm Li}_2(x)+{\rm
Li}_2({1\over 2})+{\rm Li}_2({x\over 2(x-1)})+{\rm
Li}_2(x-1)+{1\over 2}\ln^2(2-2x)$. Substituting $x=1-\beta$ and
$x=-{1-\beta\over 1+\beta}$ into this identity separately, taking
the respective difference, one then obtains the desired answer.}:
$$
{\rm Li}_2\left({1-\beta\over 2}\right) + {\rm
Li}_2\left(-{1-\beta\over 1+\beta}\right) + {1\over 2} \ln^2
\left({1+\beta\over 2} \right) =0.
$$

The occurrence of the imaginary part in (\ref{analy:f:function}) is
linked with the ``unitarity bound" of the branching ratio,  which is
obtaining from tieing the amplitudes of $\eta_Q\to \gamma\gamma$ and
$\gamma\gamma\to l^-l^+$ together according to the cutting rule.

In Nature heavy quarks are generally much heavier than leptons,
especially for the first two generations. It is then useful to know
the asymptotic behavior of the $f$ function in the $r\to 0$ limit:
\begin{subequations}
\bqa
{\rm Re}f_{\rm asym}(r)&=& {1\over 4}\ln^2 r + (1- \ln 2) \ln r +
\ln^2 2 -2 \ln 2 +{\pi^2\over 12} + {\mathcal O}(r\ln^2 r)
\,,\label{Re:f:asymptotic}
\\
{\rm Im}f_{\rm asym}(r)& = & \pi\left( {1\over 2}\ln r- \ln 2+ {\cal
O}(r\ln r) \right)\,.
\label{Im:f:asymptotic}
\eqa
\end{subequations}
As anticipated, $f(r)$ depends on $r$ only logarithmically. From
Fig.~\ref{shape:of:f}, one can see that the function $f_{\rm
asym}(r)$ already constitutes a rather good approximation as $r \le
0.1$. To reproduce these asymptotic behaviors more efficiently, one
may appeal to the {\it method of region}~\cite{Beneke:1997zp} by
dismembering the original loop integration into the sum of
integrations from different regions, e.g., hard, soft,
collinear-to-$l^-$ and collinear-to-$l^+$ in our
case~\cite{Smirnov:2001in}. One may identify the double logarithm in
(\ref{Re:f:asymptotic}) that originates
from the overlap between collinear and soft singularities.

In passing we remark on one peculiarity reported in a recent
calculation of the same process that employing the light-cone
approach~\cite{Yang:2009kq}. In that approach, the amplitude is
expressed as the convolution of a hard-scattering part with the
light-cone distribution amplitude of $\eta_Q$. The hard part there
is found to scale as $\ln r/\sqrt{r}$ in the limit $r\to 0$. This
infrared sensitivity is theoretically disastrous because it
diametrically conflicts with the requirement of helicity
suppression-- that this process should be strictly forbidden for a
massless lepton. This nuisance, if confirmed to persist, may
indicate that, the light-cone operator product expansion (OPE),
which underlies the calculational framework in \cite{Yang:2009kq},
may no longer be suited to describe heavy quarkonium decay. The
light-cone expansion is usually formulated as an expansion in terms
of a small energy scale like quark mass over a large momentum
transfer. It should be appropriate for a high-energy exclusive heavy
quarkonium production processes, as the large momentum transfer
scale can be identified with the center-of-mass energy of the
reaction, which may indeed be much greater than the heavy quark
mass. However in the process at hand, $m_Q$ itself already acts as
the highest energy scale, so it is difficult to imagine the actual
meaning of the light-cone expansion here. In a space-time picture,
the slowly-moving heavy quark and antiquark typically experience an
instantaneous strong force, so they are typically separated by a
distance of order $1/m_Q v$ but local in time, which is quite far
from a light-like separation. In our opinion, NRQCD factorization
approach, which is closely related to a local OPE by treating
$1/m_Q$ as an expansion parameter, provides the most natural and
economic framework to account for the heavy quarkonium decay, in
particular for the processes considered in this work.

It is also interesting to look at the alternative limit $r\to 1$
($\beta\to 0$), where the mass of $\eta_Q$ is just sitting at the
threshold of twice lepton mass. From (\ref{analy:f:function}), one
finds the following limiting value for $f$:
\bqa
f(1)& = & -2+2\ln 2- i\pi,
\eqa
which can also be seen in Fig.~\ref{shape:of:f}. As noted in
Ref.~\cite{Bergstrom:1982zq}, this is a well-known result, which is
responsible to the two-photon annihilation contribution to the
hyperfine splitting between the orthopositronium and
parapositronium~\cite{Karplus:1952wp} (see also
\cite{Labelle:1997uw,Pineda:1998kj}).


\subsection{The weak-interaction contribution to $\eta_Q\to l^- l^+$}

Neutral quarkonium decay is normally not a good place to look for
the trace of weak interaction, which is generally overshadowed by
the strong and electromagnetic interactions. However, for the
processes at hand, the LO QED amplitude has to proceed at one loop
order, but the weak interaction contribution instead can start at
tree level, so it is not inconceivable that the weak interaction may
play some role for some of these processes.

There is only one $s$-channel diagram that contributes to $\eta_Q
\to l^+ l^-$ from $Z^0$ exchange, as depicted in
Fig.~\ref{Feynman:diag}~$a2)$. Only the axial vector coupling of
$Z^0f\bar{f}$ contributes to this process. At the lowest order in
the velocity expansion, the weak amplitude can be expressed as
\bqa
\mathscr{M}_{\rm WEAK} &=& 2\sqrt{2N_c} \,{\psi_{\eta_Q}(0)\over
\sqrt{m_Q}} \, {  \pi \alpha \,g_A^l g_A^Q \over M_Z^2
\sin^2\theta_W \cos^2\theta_W} \,[m_l\, \bar{u}(p_1)\gamma_5
v(p_2)],
\label{WEAK:ampl:etac:l+:l-} \eqa
where $\theta_W$ is the Weinberg angle, and $g_A^l$, $g_A^Q$ denote
the weak axial charge of charged lepton and quark, respectively. The
weak axial charge of a fermion is equal to its 3rd component of weak
isospin  ($g_A^l =-{1\over 2}$ for $l=e,\mu,\tau$; $g_A^c ={1\over
2}$, $g_A^b = -{1\over 2}$). We have neglected the $\eta_Q$ mass as
well as the width of $Z^0$ in the $Z^0$ propagator, since they are
much smaller than $Z^0$ mass for $Q=c,\,b$. Therefore the $Z^0$
exchange can be effectively mimicked by a four-fermion contact
interaction. Note there is an explicit factor of $m_l$ in
Eq.~(\ref{WEAK:ampl:etac:l+:l-}), again due to helicity suppression,
very similar to what occurs to $\pi^+\to l^+\nu_l$.

Comparing (\ref{WEAK:ampl:etac:l+:l-}) with
(\ref{QED:ampl:etac:l+:l-}), one clearly sees that, though the weak
interaction amplitude is suppressed by a factor of $m_Q^2/M_Z^2$
relative to the electromagnetic one, it suffers less suppression by
one power of $\alpha$. For $\eta_b$ decay to a lepton pair, these
two competing effects may become comparable in magnitude. As a
consequence, in order to make reliable predictions, the weak
interaction effect becomes indispensable and must be included.

\subsection{The Standard Model prediction to $\eta_Q\to l^- l^+$}

Substituting Eq.~(\ref{QED:ampl:etac:l+:l-}) and
Eq.~(\ref{WEAK:ampl:etac:l+:l-}) into
\bqa
\Gamma^{\rm SM}[\eta_Q \to l^- l^+] &=& {\beta \over 16\pi
M_{\eta_Q}} \left|{\mathscr M}_{\rm EM}+{\mathscr M}_{\rm WEAK}
\right|^2,
\label{Width:formula:etac:l+:l-}
\eqa
we then get the desired partial width expected in the frame of SM.

It is convenient to introduce the {\it normalized} decay rate of
$\eta_Q \to l^- l^+$, defined as
Eq.~(\ref{Width:formula:etac:l+:l-}) normalized to the partial width
of $\eta_Q\to \gamma\gamma$:
\bqa
R^{\rm SM}[{\eta_Q\to l^+ l^-}] & \equiv& {\Gamma^{\rm SM}[\eta_Q
\to l^+ l^-] \over \Gamma_0}
\nn\\
&= & {\alpha^2 \over 2\pi^2} \, \beta \, r \left| f(r)- g_A^Q
{\sqrt{2}\, G_F M^2_{\eta_Q} \over 8 \, e_Q^2\,\alpha^2} \right|^2,
\label{R:ratio:full:etaQ:l+:l-}
\eqa
where we have substituted $e_l=-1$, and used the relation $M_Z=
({\pi\alpha\over \sqrt{2}\, G_F})^{1/2}{1\over \sin\theta_W \cos
\theta_W}$, to condense the expression for the scaled weak amplitude
($G_F$ is the Fermi coupling constant). An important feature is that
the relative importance of weak interaction contribution increases
with $M_Q$. The partial width of $\eta_Q\to \gamma\gamma$ is given
at the lowest order in $v$ and $\alpha_s$:
\bqa
\Gamma_0 &\equiv & \Gamma[\eta_Q \to \gamma\gamma] = {4 N_c \pi
e_Q^4 \alpha^2 \over m_Q^2} \psi^2_{\eta_Q}(0).
\label{width:etaQ:2photon}
\eqa
The advantage of introducing the $R$ ratio in
(\ref{R:ratio:full:etaQ:l+:l-}) is that the nonperturbative factor
$\psi_{\eta_c}(0)$ cancels out in the ratio, and some portions of
QCD radiative and relativistic corrections to both $\eta_Q\to
l^-l^+$ and $\eta_Q\to \gamma\gamma$ may largely cancel. Therefore,
for a LO calculation like this work, using the $R$ ratio instead of
the branching ratio is presumably more appropriate. In the situation
where the $\eta_Q$ di-photon decay has been experimentally measured,
one can directly obtain the $\mathcal{B}[\eta_Q \to l^+l^-]$ by
multiplying $\mathcal{B}_{\rm exp}[\eta_Q \to \gamma\gamma]$ with
the predicted $R$ ratio.

\section{Dalitz decays $\eta_Q\to l^+ l^-\gamma$}
\label{etaQ:Dalitz:decay}

It is an experimental fact that, for light pseudoscalar mesons, the
branching fractions of the Dalitz decay processes are several orders
of magnitude greater than those of the respective leptonic decays.
Certainly, it is worthwhile to examine whether the same pattern also
holds for pseudoscalar quarkonium decays or not.

We note that, an analogous pseudoscalar quarkonium strong decay
process, i.e. $\eta_Q\to q\bar{q}g$, has been previously studied by
several authors by retaining a nonzero mass for $q$. The analytic
expressions for the energy distributions of the gluon, quark and the
integrated decay rate, has been presented in
Ref.~\cite{Parkhomenko:1998kv}. In the following, we will
independently derive the corresponding energy spectra of the photon,
lepton and the integrated decay rate for the $\eta_Q$ Dalitz decay
process. When inserting a proper color factor, the exact agreement
is found between our Eqs.~(\ref{photon:spectrum:3body}),
(\ref{lepton:spectrum:3body}), (\ref{R:integrated:Dalitz}) and
Eqs.~(4), (5), (6) in Ref.~\cite{Parkhomenko:1998kv}. The process
$\eta_b\to c\bar{c}g$ has also been studied numerically in
\cite{Maltoni:2004hv,YJZhang:2007:thesis,Hao:2007rb}.

\subsection{Squared amplitude of quarkonium Dalitz decay}

In contrast to the rare decay $\eta_Q \to l^+l^-$, the Dalitz decay
$\eta_Q \to l^+l^-\gamma$ can start at tree level in QED~\footnote{
Weak interaction contribution can also start at tree level,
nonetheless is completely negligible for both $\eta_c$ and $\eta_b$
Dalitz decays.}. There are in total two diagrams, one of which is
depicted in Fig.~\ref{Feynman:diag}$b)$. The momenta of $\eta_Q$,
$l^-$, $l^+$, $\gamma$ are assigned as $P$, $p_1$, $p_2$, $p_3$,
respectively. This process first proceeds through $\eta_Q\to
\gamma\gamma^*$, and the virtual photon then fragments into a lepton
pair. If the lepton is much lighter than the quark, one expects that
the decay rate is dominated by the kinematic configuration where the
invariant mass of lepton pair is close to its minimum, $2m_l$. In
other word, it is the nonzero lepton mass that cuts off the
potential collinear singularity.

It might be worrisome that the amplitude may diverge in another
kinematic configuration, i.e., where the photon becomes very soft,
and the lepton and anti-lepton move nearly back-to-back with equal
momentum. It is well known that in $J/\psi$ decay to $l^+
l^-\gamma$, infrared divergence does arise in the long wavelength
limit of photon. This divergence is in turn canceled by including
the virtual correction to $J/\psi\to l^+ l^-$, guaranteed by the
Bloch-Nordsieck theorem~\footnote{Experimentally, $J/\psi\to l^+
l^-\gamma$ events are selected by requiring the photon energy is
greater than the characteristic detector sensitivity, say, $100$ MeV
in the Fermilab E760 experiment~\cite{Armstrong:1996hg}. Those
three-body decays with photon energy less than $100$ MeV are tagged
as the $J/\psi\to l^+ l^-$ events.}. However, such mechanism
obviously does not apply to our case to sweep the potential infrared
divergence.

A little thought reveals that Fig.~\ref{Feynman:diag}$b)$ must be
regular in the $p_3 \to 0$ limit. It is most transparent to see this
in the context of nonrelativistic effective theory. To describe an
almost on-shell quark interacting with a soft photon, one is
justified to use nonrelativistic QED (NRQED). After the incoming $Q$
emits a soft photon, it has to annihilate with $\overline{Q}$ into a
virtual photon that splits into a lepton pair. Therefore, the
emission of this soft photon has to flip the spin of $Q$, i.e.,
effectively induces a magnetic dipole transition, to convert the
$Q\overline{Q}$ pair from the initial $^1S_0$ state to a $^3S_1$
state. The effect of this soft-photon emission is accounted for by
the operator ${e e_Q\over 2 m_Q}\psi^\dagger { {\bm \sigma} \cdot {\bm
B}^{\rm em} }\psi$, where $\psi$ denotes the Pauli spinor field for
$Q$ in NRQED. The occurrence of the magnetic field strength, ${\bm
B}^{\rm em}$, which brings forth a factor of ${\bm p}_3$, will
protect against the infrared singularity arising in the quark
propagator, therefore the amplitude is infrared finite~\footnote{By the similar
reasoning, it is easy to see that the decay 
$\chi_{QJ}\to l^+l^-\gamma$ does develop an infrared
singularity in the soft-photon limit, since the corresponding 
NRQED operator governing the soft-photon emission is
of the electric dipole type, i.e. ${-i e e_Q\over 2 m_Q}\psi^\dagger \big[{{\bm D} \cdot {\bm
A}^{\rm em} }+ {{\bm
A}^{\rm em} \cdot {\bm D}}\big]\psi$, and there appears 
no factor of photon momentum to kill the corresponding one 
in the denominator.}.

At the lowest order in strong coupling and in $v$, the decay
amplitude can be routinely obtained:
\bqa
\mathscr{M}[\eta_Q \to l^- l^+ \gamma] &= & 2\sqrt{2N_c} e^3 e_Q^2
e_l {\psi_{\eta_Q}(0)\over \sqrt{m_Q}}\,
{\epsilon_{\mu\alpha\nu\beta}\, p_3^\alpha
\,\varepsilon_\gamma^{*\nu} \,P^\beta \over (P\cdot p_3)
(p_1+p_2)^2} \bar{u}(p_1)\gamma^\mu v(p_2),
\label{3body:Dalitz:amplitude}
\eqa
where $\varepsilon_\gamma$ signifies the photon polarization vector.
We have taken the undrawn diagram into account, which yields
identical contribution as Fig.~\ref{Feynman:diag}$b)$ owing to the
$C$-invariance.

From (\ref{3body:Dalitz:amplitude}) we are reassured that, the
numerator of the amplitude does contain a factor of photon momentum,
which is crucial to tame the infrared singularity arising from the
quark propagator. It is also worth noting that, since the lepton and
anti-lepton directly come from the photon fragmentation, they
necessarily form a ${}^3S_1$ state. There is no helicity suppression
mechanism affiliated with this process, consequently no factor of
$m_l$ manifests in (\ref{3body:Dalitz:amplitude}).

For notational abbreviation, we introduce three dimensionless energy
variables as $x_i = 2 E_i/ M_{\eta_Q} = 2 P\cdot p_i/ P^2$
($i=1,2,3$), where $\sqrt{P^2}= M_{\eta_Q}$, $E_i \equiv p_i^0$ is
the energy of each final state particle in the $\eta_Q$ rest frame.
We also adopt the same definition as in the previous section, $r
\equiv 4 m_l^2/ M^2_{\eta_Q}$.

It is straightforward to square the amplitude and sum over the
polarizations of final-state particles. We then obtain
\bqa
\sum |{\mathscr M}|^2 & = & {2^9 N_c \pi^3 e_Q^4 \alpha^3\over
m_Q^3}\psi^2_{\eta_Q}(0)
\nn \\
&\times& {\big[x_1(1-x_1)+x_2(1-x_2)\big]\,x_3 -
2(1-x_1)(1-x_2)+{r\over 2} x_3^2 \over x_3^2 (1-x_3)^2},
\label{squared:ampl:Dalitz}%
\eqa
which is symmetric under the interchange between $x_1$ and $x_2$. As
expected, each of the three terms in the numerator scales as $x_3^2$
in the infrared limit $x_3\to 0$.

Energy conservation demands that $x_1+x_2+x_3=2$. Any kinematic
invariants can be expressed in terms of any pair of these three
scaled energy variables. In the following we shall derive the
inclusive energy distributions of $l^-$ and $\gamma$ separately,
hence it is convenient to choose $x_1$ and $x_3$ as the two
independent variables in expressing the differential three-body
phase space. The dimensionless energy variable of $l^+$, $x_2$, has
been eliminated in the amplitude squared. The energy spectra of
$\gamma$ and $l^-$ can be obtained by choosing the different order
of two-fold integrations over $x_1$ and $x_3$. As a consequence, the
decay rate can be expressed in the following two different ways:
\begin{subequations}
\bqa
\Gamma[\eta_Q \to l^- l^+ \gamma]  &=&  { M_{\eta_Q} \over 32
(2\pi)^3} \int^{1-r}_0 d x_3 \int^{x_1^+}_{x_1^-} d x_1 \sum
|{\mathscr M}|^2
\label{3body:width:first:int:x1} \\
&=&  {M_{\eta_Q} \over 32 (2\pi)^3} \int^1_{\sqrt{r}} d x_1
\int^{x_3^+}_{x_3^-} d x_3 \sum |{\mathscr M}|^2.
\label{3body:width:first:int:x3}
\eqa
\end{subequations}
The integration boundaries for the outer-layer integrals have been
labeled explicitly; for the inner-layer integral, the upper and
lower boundaries can also be readily inferred:
\begin{subequations}
\bqa
x_1^\pm &=& {2-x_3\over 2}\pm {x_3\over 2} \sqrt{1-r-x_3\over
1-x_3},
\\
x_3^\pm &=& {2(1-x_1)\over 2- x_1 \mp \sqrt{x_1^2-r}}.
\eqa
\end{subequations}

\subsection{Inclusive photon energy spectrum}

It is straightforward to deduce the energy distribution of the
photon by integrating over the variable $x_1$ in
(\ref{3body:width:first:int:x1}):
\bqa
& & {d R[\eta_Q \to \gamma(x_3) + X] \over d x_3} \equiv {1\over
\Gamma_0}{d\Gamma[\eta_Q \to \gamma(x_3) + X ] \over d x_3}
\nn \\
& = & {\alpha\over 3\pi} {x_3 (2+r-2 x_3)\sqrt{1-r-x_3}\over
(1-x_3)^{5/2}}.
\label{photon:spectrum:3body}
\eqa
As in dealing with $\eta_Q$ decay to a lepton pair, we introduce the
scaled energy distribution of $\gamma$, $d R(x_3)/d x_3$, the
differential decay rate of $\eta_Q \to \gamma+X$ normalized with
respect to $\Gamma_0$, the partial width of $\eta_Q\to \gamma\gamma$
given in (\ref{width:etaQ:2photon}). This distribution clearly
vanishes at both the lower and upper ends of the scaled photon
energy, i.e. $x_3=0$ and $x_3=1-r$. For $r\ll 1$, the spectrum is
generally featureless and negligible in most of the region, except a
sharp peak rises near the upper end of $x_3$, which centers at
$x_3\approx 1-{\sqrt{21}+1\over 4}r\approx 1-1.40 r$, and the peak
height $\approx 0.11\alpha/r$. This clearly indicates that, the
decay rate is dominated by the kinematic configuration where the
outgoing lepton pair carries a small invariant mass, owing to the
collinear enhancement.

There is also motivation to inspect the lower end of the photon
spectrum more closely. It is an experimental fact that a realistic
electromagnetic calorimeter can detect photons only down to some
minimum limiting energy $E_{\gamma\,\rm cut}$. If the photon becomes
very soft, it will not be properly registered by the electromagnetic
calorimeter. In this respect, those $\eta_Q\to l^- l^+\gamma$ events
with a very soft photon will, from the experimental perspective,
mimic the respective $\eta_Q\to l^- l^+$ event. We learn from
(\ref{photon:spectrum:3body}) that, at small $x_3$, this
differential decay rate becomes enormously suppressed relative to
that in the $x_3 \to 1$ limit, and scales linearly with $x_3$, $\sim
{\alpha\over 3\pi}\sqrt{1-r}(2+r) x_3$. Imagine we impose a
realistic cutoff $x_{3\,cut}$ on the photon energy. Those three-body
decay events will be correctly recorded as Dalitz decay events when
the fractional photon energy is greater than $x_{3\,cut}$; in
comparison, those three-body events will be tagged as the lepton
pair events when $x_3\le x_{3\,cut}$. The $R$ value integrating over
$x_3$ from 0 to this cutoff will be approximately ${\alpha\over
3\pi}\, x^2_{3\,cut}$ for small $r$. Although this is a tiny
fraction, the chance exists that for some Dalitz decays, it might
still be much greater than the extremely small decay ratio of
$\eta_Q\to l^- l^+$ as given in (\ref{R:ratio:full:etaQ:l+:l-}). If
this is the case, what are experimentally recorded as the $\eta_Q\to
l^- l^+$ events in fact receive the bulk of contributions from the
three-body Dalitz decays. As we shall see, $\eta_c\to e^- e^+$
constitutes such a very example.

\subsection{Inclusive lepton energy spectrum and photon-to-lepton fragmentation function}

\begin{figure}[htb]
\centerline{
\includegraphics[height=10.0cm]{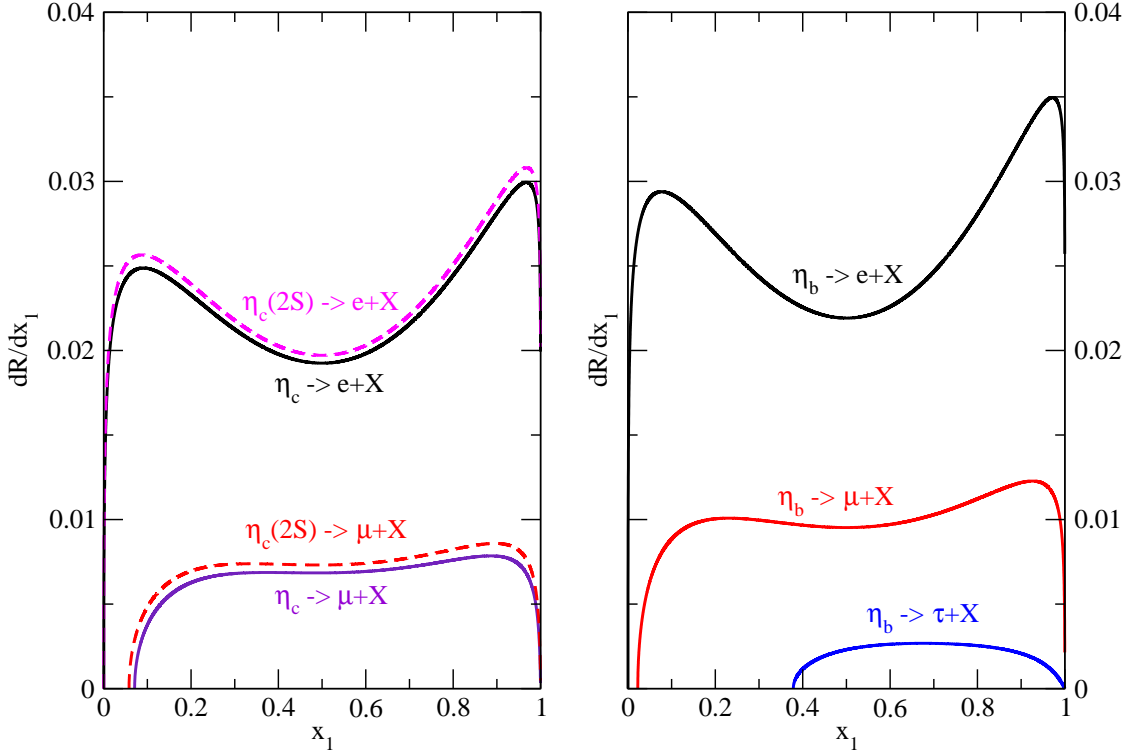}}
\caption{Normalized energy distributions of leptons in numerous
$\eta_c(\eta_c')$ and $\eta_b$ Dalitz decays. The energy spectrum of
$\tau$ in $\eta_c'$ Dalitz decay, populating a rather narrow region
near $x_1=1$, is very much suppressed with respect to the other
ones, so is not displayed in this figure.%
\label{spectrum:lepton:Dalitz} }
\end{figure}

We are also interested in the inclusive energy distribution of the
lepton. This can be obtained by integrating over the variable $x_3$
in (\ref{3body:width:first:int:x3}):
\bqa
& & {d R[\eta_Q \to l^-(x_1) + X] \over d x_1}  \equiv  {1\over
\Gamma_0}{d\Gamma[\eta_Q \to l^-(x_1) + X ] \over d x_1}
\nn\\
&= & { 2\alpha\over \pi} \left\{{x_1^2+(1-x_1)^2\over
2}\ln\left(x_1+\sqrt{x_1^2-r}\over x_1-\sqrt{x_1^2-r}\right) +
2(1-x_1)\sqrt{x_1^2-r} \right.
\nn\\
&-& \left. {1\over 2} \ln\left(2-x_1+\sqrt{x_1^2-r}\over
2-x_1-\sqrt{x_1^2-r}\right)\right\}.
\label{lepton:spectrum:3body} \eqa
As in Eq.~(\ref{photon:spectrum:3body}), we also define a
dimensionless distribution, $d R(x_1)/dx_1$, the differential decay
rate of $\eta_Q \to l^-(x_1) + X$ normalized with respect to
$\Gamma_0$. It is easy to see that this energy spectrum vanishes at
both end points: $x_1=\sqrt{r}$ and $1$. In sharp contrast to the
energy spectrum of $\gamma$, the spectrum of $l^-$ is more evenly
populated in the whole region of $x_1$. In
Fig.~\ref{spectrum:lepton:Dalitz}, we show the energy distributions
of different species of $l^-$ in various Dalitz decays of $\eta_c$,
$\eta'_c$ and $\eta_b$~\footnote{Note that the quark energy
distributions in various $\eta_Q\to q\bar{q}g$ processes, have not
been correctly displayed in Fig.~3 of \cite{Parkhomenko:1998kv}, due
to some input mistakes.}.

As $r\ll 1$, the formalism of fragmentation function can adequately
account for the leptonic energy distribution. One can readily
identify the fragmentation function $D_{\gamma\to l^-}(z)$, by
expanding (\ref{lepton:spectrum:3body}) in powers of $r$, retaining
only those nonvanishing terms in the limit $r\to 0$, and dividing
them by 2 to compensate the fact that each photon in
$\eta_Q\to\gamma\gamma$ can fragment:
\bqa
D_{\gamma\to l^-}(x_1)&= &  {\alpha\over \pi}
\left\{{x_1^2+(1-x_1)^2\over 2} \ln\left({4 x_1^2\over
r}\right)+{1\over 2}\ln(1-x_1)+2 x_1(1-x_1)\right\}.
\label{frag:func:gamma:to:lepton}
\eqa
As expected, the coefficient of the collinear logarithm is nothing
but the Altarelli-Parisi splitting kernel $P_{\gamma\to l^-}$. We
have checked that, for each Dalitz decay channel $\eta_Q\to
l^+l^-\gamma$ (except $l=\tau$), this fragmentation function
approximates the exact spectra (\ref{lepton:spectrum:3body}) quite
well, and also gives a satisfactory account of the corresponding
integrated decay rates.

\subsection{The integrated decay rate of $\eta_Q\to l^+l^-\gamma$}

It is also desirable to know the total decay rate of $\eta_Q\to
l^+l^-\gamma$, by integrating the amplitude squared in
(\ref{squared:ampl:Dalitz}) over the full three-body phase space.
The integrated decay ratio can be readily deduced by starting from
either the photon spectrum in (\ref{photon:spectrum:3body}) or the
lepton distribution in (\ref{lepton:spectrum:3body}). The final
answer admits a particularly succinct form:
\bqa
R[{\rm integrated}] & \equiv &
{\Gamma[\eta_Q \to l^-l^+\gamma] \over \Gamma_0}= {2\alpha \over 3\pi} \left\{
\ln\left(1+\sqrt{1-r}\over 1-\sqrt{1-r}\right)-{2 (4-r)\over
3}\sqrt{1-r}\right\}.
\label{R:integrated:Dalitz}
\eqa
Its limiting behavior in small $r$ reads
\bqa
R[{\rm integrated}] & = & {2\alpha \over 3\pi} \left\{\ln \left({1\over r}\right) + 2\ln
2 -{8\over 3}+ {\mathcal O}(r) \right\}.
\eqa
As expected, the greater the disparity between quark mass and lepton
mass is, the larger the integrated $R$ value becomes due to the
increasing logarithmic enhancement. This asymptotic expression can
also be easily reproduced by integrating twice of the fragmentation
function $D_{\gamma\to l^-}(x_1)$ in
(\ref{frag:func:gamma:to:lepton}) over the entire range of
the energy fraction $x_1$ of $l^-$.

It is also instructive to look at the alternative limit $r\to 1$,
where the $\eta_Q$ is barely heavy enough to disintegrate into a
leptons pair at rest plus a zero-energy photon:
\bqa
R[{\rm integrated}] & = & {4\alpha \over 15\pi}\,\left\{\beta^5+
\mathcal{O}(\beta^7)\right\},
\label{R:int:Dalitz:limit:beta:1}
\eqa
where $\beta\equiv \sqrt{1-r}$. This somewhat unexpectedly severe
suppression, will be useful for us to understand the very small
decay rate of $\eta_c(2S)\to \tau^+\tau^-\gamma$.

\section{Phenomenology}
\label{phenomenology}

In this section we will explore the consequences of the formulas
presented in previous sections. In particular, we will analyze
numerous leptonic and Dalitz decay channels of pseudoscalar
charmonia and bottomonia, to assess their observation potentials in
the current and forthcoming high-energy collision facilities.

In Table~\ref{Table:exclusive:P:l+:l-}, we tabulate numerous
predictions for $\eta_c$, $\eta_c(2S)$ and $\eta_b$ decays to all
possible species of lepton pairs. When evaluating the mass ratio
$r$, we take the precisely known lepton
masses~\cite{Amsler:2008zzb}: $m_e=0.511$ MeV, $m_\mu=105.66$ MeV,
and $m_\tau=1776.84$ MeV, and use the central values of the less
precisely measured pseudoscalar quarkonium
masses~\cite{Amsler:2008zzb}: $M_{\eta_c}=2980.3$ MeV,
$M_{\eta_c(2S)}=3637.0$ MeV, and $M_{\eta_b}=9388.9$
MeV~\cite{Aubert:2008vj}. For the electromagnetic and weak
couplings, we take $\alpha=1/137$ (for simplicity, we have neglected
the running effect of fine structure constant), and Fermi coupling
constant $G_F=1.166\times 10^{-5}$ ${\rm GeV}^{-2}$. We have taken
the electric charges of quarks to be $e_c=2/3$ and $e_b=-1/3$, the
weak axial charges of quarks to be $g_A^c=1/2$, $g_A^b=-1/2$,
respectively.

\begin{table}[t]
\caption{\label{Table:exclusive:P:l+:l-}%
The values of $r$, the scaled amplitudes for electromagnetic and
weak interactions, and the respective $R$ ratios for various
pseudoscalar quarkonium decays to a lepton pair. $R^{\rm SM}$ is the
$R$ ratio that includes both QED and weak contribution, as indicated
in (\ref{R:ratio:full:etaQ:l+:l-}); $R^{\rm EM}$ denotes the
corresponding $R$ ratio by retaining only the QED contribution. For
some of the decay channels, NRQCD predictions to $R^{\rm EM}$ have
also been given in \cite{Bergstrom:1982zq,YJZhang:2007:thesis}.}
\begin{ruledtabular}
\begin{tabular}{lccccc}
Decay modes & \multicolumn{1}{c}{ $r$ }& \multicolumn{1}{c} {$f(r)$}
& \multicolumn{1}{c}{$-g_A^Q {\sqrt{2} G_F M^2_{\eta_Q} \over 8\,
e_Q^2\,\alpha^2}$} & \multicolumn{1}{c}{$R^{\rm EM}$} &
\multicolumn{1}{c}{$R^{\rm SM}$}
\\
\hline
$\eta_c\to e^+ e^-$   &  $1.18\times 10^{-7}$ & $58.67-27.24 i$ &
$-0.39$ & $1.33\times 10^{-9}$ & $1.31\times 10^{-9}$
\\
$\eta_c\to \mu^+ \mu^-$ &  $5.03\times 10^{-3}$ & $5.30-10.51 i $ &
$-0.39$ & $1.88\times 10^{-6}$ & $1.82\times 10^{-6}$
\\
\hline
$\eta_c(2S)\to e^+ e^-$    &  $7.90\times 10^{-8}$ & $61.76-27.87 i$
& $-0.58$ &  $9.79\times 10^{-10}$ & $9.63\times 10^{-10}$
\\
$\eta_c(2S)\to \mu^+ \mu^-$  &  $3.38\times 10^{-3}$ & $6.27-11.13 i
$ & $-0.58$ & $1.49\times 10^{-6}$ & $1.42\times 10^{-6}$
\\
$\eta_c(2S) \to \tau^+ \tau^-$ &  $0.955$ & $-0.61-3.19 i $ &
$-0.58$ & $5.78\times 10^{-6}$ & $6.35\times 10^{-6}$
\\
\hline
$\eta_b\to e^+ e^-$ &  $1.19\times 10^{-8}$ & $77.59-30.85 i$ &
$15.35$ & $2.23 \times 10^{-10}$ & $3.07\times 10^{-10}$
\\
$\eta_b\to \mu^+ \mu^-$   &  $5.07\times 10^{-4}$ & $11.98-14.10 i $
& $15.35$ & $4.68\times 10^{-7}$ & $1.29\times 10^{-6}$
\\
$\eta_b \to \tau^+ \tau^-$  &  $0.143$ & $0.19-5.52 i $ & $15.35$ &
$1.09\times 10^{-5}$ & $9.74\times 10^{-5}$
\end{tabular}
\end{ruledtabular}
\end{table}

Listed in Table~\ref{Table:exclusive:P:l+:l-} are various normalized
$R$ values. Since each leptonic decay of $\eta_Q$ is severely
suppressed with respect to $\eta_Q \to \gamma\gamma$,  $R\ll 1$
certainly is expected. If one wishes to convert these $R$ ratios to
the corresponding branching fractions, one should multiply them by
the respective diphoton branching fraction $\mathcal{B}[\eta_Q \to
\gamma\gamma]$. For $\eta_c$, we may use the measured result
$\mathcal{B}_{\rm exp}[\eta_c \to
\gamma\gamma]=2.4^{+1.1}_{-0.9}\times
10^{-4}$~\cite{Amsler:2008zzb}. For $\eta_c(2S)$, only an upper
bound has been set experimentally, $\mathcal{B}_{\rm
exp}[\eta_c^\prime\to\gamma\gamma]<5\times 10^{-4}$. However, there
is good reason to believe
$\mathcal{B}[\eta_c(2S)\to\gamma\gamma]\approx
\mathcal{B}[\eta_c\to\gamma\gamma]$. To date the decay $\eta_b
\to\gamma\gamma$ has not yet been observed experimentally.
Theoretically, the branching fraction of $\eta_Q\to \gamma\gamma$
can be estimated in NRQCD factorization
approach~\cite{Bodwin:1994jh}~\footnote{ 
Note that this formula works poorly for $\eta_c$ decay.
For a more satisfactory estimate of $\mathcal{B}[\eta_c\to \gamma\gamma]$,
it is important to resum a class of contributions associated with the
running of the strong coupling $\alpha_s$~\cite{Bodwin:2001pt}.}:
\bqa
&& {\mathcal{B}[\eta_Q \to \gamma\gamma]} \approx  {\Gamma[\eta_Q
\to \gamma\gamma]\over \Gamma[\eta_Q \to g g]}
\nn\\
 &\approx & {9\, e_Q^4 \alpha^2 \over 2\, \alpha_s^2(2m_Q)}
 {1+({\pi^2\over 4}-5)C_F{\alpha_s(2m_Q)\over \pi}
 \over 1+ \big[({\pi^2\over 4}-5) C_F
 + ({199\over 18}-{13\pi^2\over 24})C_A-
 {8\over 9}n_f \big] {\alpha_s(2m_Q)\over \pi}}\,,
\label{branching:ratio:etaQ:2photon}
\eqa
where we have approximated the total hadronic width of $\eta_Q$ by
its gluonic width, and have included the NLO QCD corrections for
both $\eta_Q\to \gamma\gamma$ and $\eta_Q\to gg$. $C_F={N_c^2-1\over
2N_c}$, $C_A=N_c$, are the Casmirs for the fundamental and adjoint
representations of $SU(3)_c$ group, respectively. Taking the number
of active light flavors $n_f=4$ for $\eta_b$, and
$\alpha_s(2m_b)=0.18$, we then get $\mathcal{B}[\eta_b \to
\gamma\gamma] \approx 4.7 \times 10^{-5}$.

Owing to the energy conservation, $\eta_c$ can only decay to
$e^+e^-$ and $\mu^+\mu^-$, while $\eta_c(2S)$ and $\eta_b$ can
access all three generations of leptons. From
Table~\ref{Table:exclusive:P:l+:l-}, one can see that the values of
$r$ span a rather wide range, from the smallest $10^{-8}$ in
$\eta_b\to e^+ e^-$ to the largest $0.96$ in $\eta_c'\to \tau^+
\tau^-$. Therefore, quarkonium decays to a lepton pair seem to
provide a richer theoretical playground than the analogous decays of
light pseudoscalar mesons.

In the following we summarize the main lessons we have learnt from
Table~\ref{Table:exclusive:P:l+:l-}:
\begin{enumerate}
\item{
Among all the studied pseudoscalar quarkonium decays to lepton pair,
$\eta_b\to \tau^+\tau^-$ seems to have the largest branching ratio,
$\approx 5\times 10^{-9}$. However, the number of produced $\eta_b$
at Super $B$ experiment may not be copious enough for observing this
decay mode. On the other hand, it also looks rather challenging to
tag this decay mode in the high-energy hadron collider experiment
such as LHC, since the $\tau$ events are difficult to reconstruct in
hadronic collision environment. It is also interesting to note that,
the decay $\eta_b\to \mu^+\mu^-$, with a branching ratio as small as
$10^{-10}$, seems comparable in cleanness with the decay chain
$\eta_b\to J/\psi J/\psi\to 4\mu$, the ``golden mode" for hunting
$\eta_b$ at LHC, which has an estimated branching ratio of
$(0.7-6.7)\times 10^{-10}$~\cite{Gong:2008ue}. It may be worthwhile
to look for this dimuon decay mode at LHC, but likely the signal
events would be completely swallowed by the copious backgrounds. The
rarest decay channels, $\eta_b \to e^+e^-$, $\eta_c (\eta_c') \to
e^+e^-$, with branching ratios about $10^{-14}-10^{-13}$, seems
completely out of the reach of any foreseeable experiments.
}

\item{
For light pseudoscalar mesons decays to a lepton pair, one often
resorts to the {\it unitarity bound}, which is 
obtained from cutting the intermediate photon lines in the amplitude,
to estimate the decay rate of
$P\to l^+l^-$. In some cases this simplified but model-independent
predictions seems not far below the exact results. Inspecting the
phase pattern of $f(r)$ in Table~\ref{Table:exclusive:P:l+:l-}, it
is clear that this approximation can not make an accurate account
for the majority of pseudoscalar quarkonia leptonic decay processes.
}

\item{
The helicity suppression mechanism, which is manifested in the
prefactor $r$ in (\ref{R:ratio:full:etaQ:l+:l-}), plays a prominent
role in dictating the size of each leptonic decay rate, and is much
more important than the logarithmically running $f$ function. This
is clearly seen in the smallness of the $R$ ratio for $\eta_b \to
e^+e$, even though the respective $|f|$ is the largest among all the
decay channels. In an alternative case, $\eta_c(2S)\to
\tau^+\tau^-$, which hardly suffers from helicity suppression
because of the rather large $r$, the respective branching ratio
nevertheless remains small. This may be partly ascribable to the
small $|f|$, and partly to the rather limited phase space available
for this process, recalling the factor $\beta$ contained in
(\ref{R:ratio:full:etaQ:l+:l-}).
}

\item{
The contribution of weak interaction is insignificant in
$\eta_c(\eta'_c)$ decay, but can become important in $\eta_b$
leptonic decay. For example, including the $Z^0$ exchange effect
will significantly enhance the branching fraction of $\eta_b\to
\tau^+\tau^-$ predicted by QED alone, almost by one order of
magnitude! This can be clearly understood from the fact that, the
relative importance of the weak interaction effect grows with $m_Q$
(see (\ref{R:ratio:full:etaQ:l+:l-})). The interference between QED
and weak interaction can be either destructive or constructive,
depending on the sign of the axial charges of heavy quarks, and also
on the sign of the real part of the $f$ function.
}
\end{enumerate}

\begin{table}[t]
\caption{\label{Table:Dalitz:P:l+:l-}%
The normalized $R$ ratios for various pseudoscalar quarkonia Dalitz
decays. $R[E_{\gamma}<20\;{\rm MeV}]$ represent the corresponding
ratio with a 20 MeV cutoff imposed on the photon energy, so these
Daltiz events may be experimentally indistinguishable from those
$\eta_Q \to l^+l^-$ events. $R[{\rm integrated}]$ is the $R$ value
integrated over the whole three-body phase space. For comparison, we
also juxtapose the SM predictions to the $R$ values for pseudoscalar
quarkonia decays to a leptonic pair, which are lifted from
Table~\ref{Table:exclusive:P:l+:l-}.}
\begin{ruledtabular}
\begin{tabular}{lcccc}
Decay modes &    \multicolumn{1}{c}{ $r$ } &
\multicolumn{1}{c}{$R^{\rm SM}[{\eta_Q\to l^+l^-}]$} &
\multicolumn{1}{c}{$R[E_\gamma < 20\;{\rm MeV}]$} &
\multicolumn{1}{c}{$R[{\rm integrated}]$}
\\
\hline
$\eta_c\to e^+ e^-\gamma$  &  $1.18\times 10^{-7}$ & $1.31\times
10^{-9}$ & $1.41\times 10^{-7}$ & $0.0227$
\\
$\eta_c\to \mu^+ \mu^-\gamma$ &  $5.03\times 10^{-3}$ & $1.82 \times
10^{-6}$ & $1.41\times 10^{-7}$& $0.0062$
\\
\hline
$\eta_c(2S)\to e^+ e^-\gamma$ &  $7.90\times 10^{-8}$ & $9.63\times
10^{-10}$ & $9.44\times 10^{-8}$ & $0.0233$
\\
$\eta_c(2S)\to \mu^+ \mu^-\gamma$  &  $3.38\times 10^{-3}$ &
$1.42\times 10^{-6}$ & $9.44\times 10^{-8}$ & $0.0068$
\\
$\eta_c(2S) \to \tau^+ \tau^-\gamma$  &  $0.955$ & $6.35\times
10^{-6}$ & $2.73\times 10^{-8}$ & $2.8\times 10^{-7}$
\\
\hline
$\eta_b\to e^+ e^-\gamma$ & $1.19\times 10^{-8}$ & $3.07 \times
10^{-10}$ & $1.41\times 10^{-8}$ & 0.0263
\\
$\eta_b\to \mu^+ \mu^-\gamma$ & $5.07\times 10^{-4}$ & $1.29\times
10^{-6}$ & $1.41\times 10^{-8}$ & 0.0098
\\
$\eta_b \to \tau^+ \tau^-\gamma$ & $0.143$ & $9.74 \times 10^{-5}$ &
$1.40\times 10^{-8}$ & 0.0014
\end{tabular}
\end{ruledtabular}
\end{table}

Next we explore the phenomenological consequences of the Dalitz
decay $\eta_Q\to l^+l^-\gamma$. In Table~\ref{Table:Dalitz:P:l+:l-}
we tabulate the $R$ ratios integrated over the full three-body phase
space, as well as the corresponding $R$ ratios by integrating over
the photon energy up to 20 MeV. The purpose of including the latter
is to assess the likelihood for these Dalitz events to fake the
exclusive $\eta_Q\to l^+l^-$ event. The main understanding we gained
from Table~\ref{Table:Dalitz:P:l+:l-} are:
\begin{enumerate}
\item{
These three-body Dalitz decay processes in general have branching
ratios several orders of magnitude larger than the corresponding
$\eta_Q\to l^+l^-$ decay. This drastic disparity should be
attributed to several factors: less suppression by powers of
$\alpha$, the absence of helicity suppression, and the collinear
enhancement. As a result, the branching ratios of
$\eta_c(\eta'_c)\to e^+ e^-\gamma$ may reach $4\times 10^{-6}$, 
and those of $\eta_c(\eta'_c)\to
\mu^+ \mu^-\gamma$ may reach $10^{-6}$. 
In the recently launched BESIII experiment, roughly
$10^{10}$ $J/\psi$ events are expected to be accumulated. $\eta_c$
can be most copiously produced from $J/\psi$ through the magnetic
dipole transition, with $\mathcal{B}_{\rm
exp}[J/\psi\to\eta_c\gamma]=1.3\pm0.4\%$~\cite{Amsler:2008zzb}.
Therefore about $10^8$ $\eta_c$ events are expected to be collected,
and about ${\mathcal O}(10^2)$ Dalitz decay events should be
produced. Even taking the detection acceptance and efficiency into
account, the observation prospect for the aforementioned Dalitz
decays at BES-III still looks optimistic~\footnote{However, the
decay $\eta_c \to e^+ e^-\gamma$ at BES-III experiment may be
subject to substantial contamination from the background Bhabha
events. In comparison, the decay $\eta_c\to \mu^+ \mu^-\gamma$ might
be easier to establish.}. Similarly, if the future Super $B$ factory
can accumulate a huge $\eta_b$ samples, it may again be feasible to
look for the $\eta_b \to e^+ e^-\gamma,\, \mu^+ \mu^-\gamma$ events.
It is obvious that the majority of Dalitz decay events will obey a
fragmentation pattern, i.e., a hard photon of $E_\gamma \approx
{M_{\eta_Q}\over 2}$ recoiling against a pair of nearly collinear
leptons.
}

\item{
$\eta_c(2S)\to \tau^+\tau^-\gamma$ is the only exceptional Dalitz
decay process that has a even smaller decay rate than its respective
leptonic decay $\eta_c(2S)\to \tau^+\tau^-$. It is interesting to
trace the reason. First by recalling
(\ref{R:int:Dalitz:limit:beta:1}), we note the rather strong
suppression of the integrated decay ratio of the Dalitz decay near
the mass threshold, $\propto \beta^5$. By contrast, the suppression
of the decay $\eta_c(2S)\to \tau^+\tau^-$ is only linear in $\beta$,
stemming from the two-body phase space. It is these two very
different threshold scaling behaviors that result in this anomalous
pattern.
}

\item{
Suppose a realistic electromagnetic calorimeter, say, the one
installed in the BES-III detector, can detect those photons only
with energy greater than 20 MeV~\cite{Asner:2008nq}. When such a
cutoff is imposed, according to Table~\ref{Table:Dalitz:P:l+:l-},
the number of Dalitz decay events from $\eta_c(\eta'_c) \to e^+
e^-\gamma$ with photon energy less than this cutoff, turns out to be
about $100$ times greater than that from $\eta_c(\eta'_c) \to e^+
e^-$. Therefore, what are experimentally recorded as the
$\eta_c(\eta'_c) \to e^+ e^-$ events, in fact receive the bulk of
contribution from the three-body Dalitz decay events with
unregistered soft photons. However, even if incorporating this two
orders-of-magnitude enhancement, the corresponding decay
probabilities are still too low for such channels to be established
at BES-III experiment.
A valuable lesson learned from this example is that, in any attempt
to interpret the possible $\eta_Q\to l^+l^-$ event as the signature
for new interactions beyond SM, one must ensure that the
contamination from the corresponding Dalitz decays has already been
thoroughly understood and carefully incorporated in the analysis.
}
\end{enumerate}

\section{Summary and Outlook}
\label{summary}

In this work, we have performed a comprehensive analysis of
pseudoscalar quarkonium decays to a lepton pair without and with
bremsstrahlung. These rare electromagnetic decay processes offer a
clean platform to test our understanding of quarkonium dynamics. We
corroborate the previous conclusion that the exclusive decays to
lepton pair are extremely suppressed in Standard Model.
By contrast, the pseudoscalar quarkonium Dalitz decays in general
have a much larger decay rate, because of several joint factors:
less suppression by power of $\alpha$, absence of helicity
suppression, and collinear enhancement. It is found that the Dalitz
decays $\eta_c\to e^+ e^-\gamma$ and $\eta_c \to \mu^+ \mu^-\gamma$,
with branching fractions of order $10^{-6}$, may have the bright
prospect to be established in the BES-III experiment.

It is stressed that a realistic electromagnetic calorimeter can
detect photons only down to a minimum energy. Thus from the
experimental perspective, those Dalitz decay events with photon
energy less than this minimum energy, will be tagged as the
exclusive lepton pair events. Taking this fact into consideration,
it is found that the {\it measured} decay rate of $\eta_c(\eta'_c)
\to e^+ e^-$ would be about two orders of magnitude greater than
that {\it literally predicted} from (\ref{R:ratio:full:etaQ:l+:l-}).
Nevertheless, such amplification is still not dramatic enough to
warrant their observation at BES-III experiment. In general, the
observation prospect for the pseudoscalar quarkonium decays to a
lepton pair seems extremely pessimistic within the frame of the
Standard Model. In this respect, future unambiguous sighting of any
of this type of decays may be viewed as the strong evidence for the
existence of new physics.

Our analysis is based on a leading order calculation both in strong
coupling and in quark velocity. To improve the reliability of our
predictions, it is worthwhile to implement the QCD perturbative and
relativistic corrections to the pseudoscalar quarkonium decay
processes considered in this work, in particular to the quarkonium
Dalitz decay processes.

An interesting extension of this work is to study ${}^3 P_J$
($J=0,1,2$) quarkonium states decays to a lepton pair with and
without bremsstrahlung. The processes $\chi_{c1,2}\to e^+e^-$ and
$e^+e^-\to \chi_{c1,2}$ have been studied long ago within the
color-singlet model~\cite{Kuhn:1979bb}~\footnote{For the calculations of
$\chi_{c1,2}\to e^+e^-$, one is allowed to put $m_e$ to 0 since
these processes are not subject to the helicity suppression, and
retaining a nonzero $m_e$ only yields a small correction.}, where 
infrared divergences are reported and subsequently cured by 
imposing a phenomenological cutoff of ``binding energy". 
To our knowledge, the process $\chi_{cJ}\to l^+l^-\gamma$ 
has not been considered before, which is also plagued with infrared divergences. 
It is theoretically interesting to investigate these two processes in the context of
NRQCD (NRQED), which provide a systematic way to tame these
infrared divergences by incorporating the effect of higher Fock
state ($|c\bar{c}({}^3S_1)\gamma\rangle$ in our case, 
with the dynamical photon understood to be {\it ultrasoft}). 
We note that the latter process is quite analogous to the $\chi_{bJ}$
inclusive decays to charmed hadrons, which have recently been analyzed 
in the NRQCD factorization approach~\cite{Bodwin:2007zf}. Thus one may
simply use their results with some slight
modification. 

Eliminating the infrared divergences associated with the former 
process turns out to be more subtle, and, more interesting, 
but likely to be achievable provided that one
appeals to the even lower-energy effective theory of NRQCD, the so-called
potential NRQCD (pNRQCD)~\cite{Brambilla:1999xf}, 
by retaining the ultrasoft gluons (photons) as 
the manifest degree of freedom~\footnote{
The symptom encountered here seems to be of the
similar origin as what was found in the exclusive $B$-meson decays to $P$-wave
charmonium, {\it e.g.} $B\to \chi_{cJ}K$~\cite{Song:2003yc}. In that case,
it has been recently shown that under certain condition, 
the factorization can be recovered provided that 
the color-octet contribution is included at the amplitude level~\cite{Beneke:2008pi}.}
(for a pNRQCD-based study of the color-octet effect in exclusive reaction involving quarkonium,
such as $J/\psi\to \eta_c\gamma$, see \cite{Brambilla:2005zw}).
We hope future studies of these two types of processes
at BES-III experiment, especially the resonant production process
$e^+e^-\to \chi_{c1,2}$, may lend some important guidance.

\acknowledgments

We thank Yu-Qi Chen, Rong-Gang Ping and Yu-Jie Zhang for valuable
discussions.
We thank Alexander Parkhomenko for bringing
Ref.~\cite{Parkhomenko:1998kv} to our attention.
This work was supported in part by the National Natural Science
Foundation of China under grants No.~10875130 and No.~10875156.



\begin{thebibliography}{99}
\bibitem{Drell:1959}
  S.~D.~Drell,
  Nuovo.\ Cim.\  {\bf XI}, 693 (1959).
\bibitem{Berman:1960}
  S.~M.~Berman and D.~A.~Geffan,
  Nuovo.\ Cim.\ {\bf XVIII}, 1192 (1960).
\bibitem{Young:1967zz}
  B.~L.~Young,
  Phys.\ Rev.\  {\bf 161}, 1620 (1967).
\bibitem{Bergstrom:1982zq}
  L.~Bergstrom,
  Z.\ Phys.\  C {\bf 14}, 129 (1982).
\bibitem{Bergstrom:1983ay}
  L.~Bergstrom, E.~Masso, L.~Ametller and A.~Bramon,
  Phys.\ Lett.\  B {\bf 126}, 117 (1983).
\bibitem{Savage:1992ac}
  M.~J.~Savage, M.~E.~Luke and M.~B.~Wise,
  Phys.\ Lett.\  B {\bf 291}, 481 (1992)
  [arXiv:hep-ph/9207233].
\bibitem{Ametller:1993we}
  L.~Ametller, A.~Bramon and E.~Masso,
  Phys.\ Rev.\  D {\bf 48}, 3388 (1993)
  [arXiv:hep-ph/9302304].
\bibitem{GomezDumm:1998gw}
  D.~Gomez Dumm and A.~Pich,
  Phys.\ Rev.\ Lett.\  {\bf 80}, 4633 (1998)
  [arXiv:hep-ph/9801298].
\bibitem{Dorokhov:2007bd}
  A.~E.~Dorokhov and M.~A.~Ivanov,
  Phys.\ Rev.\  D {\bf 75}, 114007 (2007)
  [arXiv:0704.3498 [hep-ph]].
\bibitem{Dalitz:1951aj}
  R.~H.~Dalitz,
  Proc.\ Phys.\ Soc.\  A {\bf 64}, 667 (1951).
\bibitem{Amsler:2008zzb}
  C.~Amsler {\it et al.}  [Particle Data Group],
  Phys.\ Lett.\  B {\bf 667}, 1 (2008).
\bibitem{Abouzaid:2006kk}
  E.~Abouzaid {\it et al.}  [KTeV Collaboration],
  Phys.\ Rev.\  D {\bf 75}, 012004 (2007)
  [arXiv:hep-ex/0610072].
\bibitem{Asner:2008nq}
  D.~M.~Asner {\it et al.},
  arXiv:0809.1869 [hep-ex].
\bibitem{Bodwin:1994jh}
  G.~T.~Bodwin, E.~Braaten and G.~P.~Lepage,
  Phys.\ Rev.\  D {\bf 51}, 1125 (1995)
  [Erratum-ibid.\  D {\bf 55}, 5853 (1997)].
\bibitem{YJZhang:2007:thesis}
Y.~J.~Zhang, PhD thesis, {\it The study of $J/\psi$ production in
$e^+e^-$ annhilation and bottomonium decay to charm quark pair},
Peking U., 2007.
\bibitem{Yang:2009kq}
  M.~Z.~Yang,
   Phys.\ Rev.\  D {\bf 79}, 074026 (2009)
  [arXiv:0902.1295 [hep-ph]].
\bibitem{Bergstrom:1980fr}
  L.~Bergstrom and H.~Snellman,
  Z.\ Phys.\  C {\bf 8}, 363 (1981).
\bibitem{DiSalvo:2000ec}
  E.~Di Salvo, M.~P.~Rekalo and E.~Tomasi-Gustafsson,
  Eur.\ Phys.\ J.\  C {\bf 16}, 295 (2000)
  [arXiv:hep-ph/0004112].
\bibitem{Isgur:1975ib}
  N.~Isgur,
  Phys.\ Rev.\  D {\bf 12}, 3770 (1975).
\bibitem{Beneke:1997zp}
  M.~Beneke and V.~A.~Smirnov,
  Nucl.\ Phys.\  B {\bf 522}, 321 (1998).
\bibitem{Smirnov:2001in}
  V.~A.~Smirnov,
in {\it Proc. of the 5th International Symposium on Radiative
Corrections (RADCOR 2000)} ed. Howard E. Haber,
  arXiv:hep-ph/0101152.
\bibitem{Karplus:1952wp}
  R.~Karplus and A.~Klein,
  Phys.\ Rev.\  {\bf 87}, 848 (1952).
\bibitem{Labelle:1997uw}
  P.~Labelle, S.~M.~Zebarjad and C.~P.~Burgess,
  Phys.\ Rev.\  D {\bf 56}, 8053 (1997)
  [arXiv:hep-ph/9706449].
\bibitem{Pineda:1998kj}
  A.~Pineda and J.~Soto,
  Phys.\ Rev.\  D {\bf 58}, 114011 (1998)
  [arXiv:hep-ph/9802365].
\bibitem{Parkhomenko:1998kv}
A.~Y.~Parkhomenko and A.~D.~Smirnov,
Mod.\ Phys.\ Lett.\ A {\bf 13}, 2199 (1998) [arXiv:hep-ph/9808363].
\bibitem{Maltoni:2004hv}
  F.~Maltoni and A.~D.~Polosa,
  Phys.\ Rev.\  D {\bf 70}, 054014 (2004)
  [arXiv:hep-ph/0405082].
\bibitem{Hao:2007rb}
  G.~Hao, C.~F.~Qiao and P.~Sun,
  Phys.\ Rev.\  D {\bf 76}, 125013 (2007)
  [arXiv:0710.3339 [hep-ph]].
\bibitem{Armstrong:1996hg}
  T.~Armstrong {\it et al.}  [Fermilab E760 Collaboration],
  Phys.\ Rev.\  D {\bf 54}, 7067 (1996).
\bibitem{Aubert:2008vj}
  B.~Aubert {\it et al.}  [BABAR Collaboration],
  Phys.\ Rev.\ Lett.\  {\bf 101}, 071801 (2008)
  [Erratum-ibid.\  {\bf 102}, 029901 (2009)]
  [arXiv:0807.1086 [hep-ex]].
\bibitem{Bodwin:2001pt}
  G.~T.~Bodwin and Y.~Q.~Chen,
  Phys.\ Rev.\  D {\bf 64}, 114008 (2001)
  [arXiv:hep-ph/0106095].
\bibitem{Gong:2008ue}
  B.~Gong, Y.~Jia and J.~X.~Wang,
  Phys.\ Lett.\  B {\bf 670}, 350 (2009)
  [arXiv:0808.1034 [hep-ph]].
\bibitem{Kuhn:1979bb}
  J.~H.~Kuhn, J.~Kaplan and E.~G.~O.~Safiani,
  Nucl.\ Phys.\  B {\bf 157}, 125 (1979).
\bibitem{Bodwin:2007zf}
  G.~T.~Bodwin, E.~Braaten, D.~Kang and J.~Lee,
  Phys.\ Rev.\  D {\bf 76}, 054001 (2007)
  [arXiv:0704.2599 [hep-ph]].
\bibitem{Brambilla:1999xf}
  N.~Brambilla, A.~Pineda, J.~Soto and A.~Vairo,
  Nucl.\ Phys.\  B {\bf 566}, 275 (2000)
  [arXiv:hep-ph/9907240].
\bibitem{Song:2003yc}
  Z.~Z.~Song, C.~Meng, Y.~J.~Gao and K.~T.~Chao,
  Phys.\ Rev.\  D {\bf 69}, 054009 (2004)
  [arXiv:hep-ph/0309105].
\bibitem{Beneke:2008pi}
  M.~Beneke and L.~Vernazza,
  Nucl.\ Phys.\  B {\bf 811}, 155 (2009)
  [arXiv:0810.3575 [hep-ph]].
\bibitem{Brambilla:2005zw}
  N.~Brambilla, Y.~Jia and A.~Vairo,
  Phys.\ Rev.\  D {\bf 73}, 054005 (2006)
  [arXiv:hep-ph/0512369].



\end{thebibliography}
\end{document}